\documentclass[twocolumn,english,aps,prb,floatfix,comment]{revtex4}
\usepackage[T1]{fontenc}
\usepackage[latin2]{inputenc}
\usepackage{amssymb}

\makeatletter

\usepackage{graphicx}
\usepackage{bm}

\usepackage{babel}
\makeatother
\begin{document}

\title{Comment on ''Kondo resonances and Fano antiresonances in transport
through quantum dots''}

\author{J. Bon\v{c}a, A. Ram\v{s}ak, and T. Rejec}

\affiliation{J. Stefan Institute, Jamova 39, 1000 Ljubljana, Slovenia\\
 Faculty of Mathematics and Physics, University of Ljubljana, Jadranska
19, 1000 Ljubljana, Slovenia}

\begin{abstract}
We show that the numerical method {[}M.~E.~Torio, K.~Hallberg,
A.~H.~Ceccatto, and C.~R.~Proetto, Phys. Rev. B \textbf{65}, 085302
(2002){]} does not reproduce correctly the Kondo physics in quantum
dot systems with interaction. 
\end{abstract}


\date{19 July 2004}

\maketitle
In a recent paper, M.~E.~Torio, K.~Hallberg, A.~H.~Ceccatto,
and C.~R.~Proetto (THCP) presented a numerical calculation of the
conductance of an interacting system consisting of a quantum dot coupled
to the leads.\cite{torio02} Their results exhibit a fair qualitative
agreement with the corresponding experiments by J.~Göres \emph{et
al.\cite{gores00}} However, a closer quantitative inspection of the
results concerning the Kondo regime clearly shows that their method
fails to reproduce adequately the main features of the conductance
in this regime.

Numerical method of THCP consists of exact diagonalization of a small
cluster containing the region with interaction, coupling the cluster
to external reservoirs (leads), and finally self-consistent calculation
of the corresponding Dyson equation. As pointed out in THCP, the method
should be reliable as long as the corresponding Kondo cloud does not
exceed the size of the cluster. In order to meet these criteria, THCP
chose $t'/t=1/\sqrt{2}$ and estimated (without giving arguments)
the corresponding Kondo correlation length of about 10 lattice sites
for $U/\Delta=12$ with $\Delta=2t'^{2}/t$. Using this set of parameters
for the Anderson model\cite{parameters} they calculated the conductance
of substitutional quantum dot configuration. For the symmetric situation,
$\epsilon_{0}=-U/2$, their conductance reaches the unitary limit,
$G=G_{0}=2e^{2}/h$. They further claimed that this is a nontrivial
check of the method which proves that their finite-system approach
produces a Kondo peak with the exact spectral weight at the Fermi
energy.\cite{torio02}

In this Comment we show that although the conductance of THCP reaches
the unitary limit, the method does not reproduce correctly the Kondo
peak. In Fig.~1(a) the conductance of THCP is shown compared to the
result obtained with the two-point sine conductance formula from Ref.~\onlinecite{rejec03}
using the variational wave function of Gunnarsson and Schönhammer\cite{schonhammer85}
together with the constrained path quantum Monte Carlo method (CPMC).\cite{gubernatis,bonca04}
In the empty orbital regime, $|\epsilon_{0}+U/2|\gtrsim U/2$, all
results match perfectly. However, for $|\epsilon_{0}+U/2|\lesssim U/2$,
except in the symmetric point, the result of THCP drastically fails
to reproduce the main fingerprint of Kondo physics, i.e., the plateau
in the conductance in the Kondo regime. \textit{\emph{The reason is
the low energy resolution which in turn is a fingerprint of finite-size
effects of a small clusters used by the method of THCP}}. It should
be pointed out that even relatively large $t'$ cannot remedy this
deficiency of the method, because the problem falls into the unusually
\emph{narrow} bandwidth regime $2t\ll U$, and $\epsilon_{0}\ll-2t$,
$\epsilon_{0}+U\gg2t$, where the scattering processes which involve
electrons or holes in the band edges cannot cause real charge fluctuations
of the impurity,\cite{jefferson77} resulting in the reduction of
the Kondo temperature. \textit{\emph{On the one hand, narrow bandwidth
should improve the energy resolution of the method,\cite{torio02}
on the other hand, however, the Kondo temperature}} $T_{K}$ \textit{\emph{in
this limit is reduced,}} \cite{hewson93}

\begin{equation}
T_{K}\sim2t\left(\frac{\Delta U}{\mid\epsilon_{0}\mid\mid\epsilon_{0}+U\mid}\right)^{1/2}e^{-\pi\mid\epsilon_{0}\mid\mid\epsilon_{0}+U\mid/(2\Delta U)}.\label{eq:tkondo}\end{equation}
 Smaller $t$ (or larger $t'$) thus cannot bring the method \textit{\emph{of}}
\textit{}\textit{\emph{THCP}}  into the regime of reliability unless
the number of sites in the cluster is significantly increased or the
strength of the interaction, $U/\Delta$, is reduced.

In a more common limit of a wide bandwidth, $2t\gg U$,\cite{haldane78}
the conductance can be calculated exactly using the Friedel sum rule
and the Bethe ansatz result.\cite{wiegman80} In Fig.~1(b) the exact
conductance for wide bandwidth is presented, together with the result
of the two-point sine formula, using  \textit{\emph{variational wavefunction
and the CPMC method}}\emph{.} The results agree for all regimes. Small
discrepancies can be attributed to the finite bandwidth of both numerical
methods ($t=1.04U$) and to the variational nature of the wave functions
used. \textit{\emph{We would like to stress that due to the lower
Kondo temperature in the narrow bandwidth regime, Fig.~1(a), the
Kondo plateau is obviously more pronounced compared to the wide bandwidth
limit, Fig.~1(b).}}

In Fig.~1(b) the conductance using an elementary Hartree-Fock calculation
is \textit{\emph{also}} presented. It is well known that this approximation
cannot adequately reproduce the Kondo phenomenon. \textit{\emph{Nevertheless,
the unitary limit in the symmetric point,}} \textit{$\epsilon_{0}+U/2=0$,}
\textit{\emph{is in this approach as well reproduced. It is a consequence
of the particle-hole symmetry of the Anderson model combined with
the Friedel sum rule}}\emph{.} The existence of the unitary limit
can in general serve only as a test of internal numerical consistency
of a particular numerical method and is not a direct evidence of a
correct description of the Kondo physics as argued by THCP.

\widetext

\begin{figure}
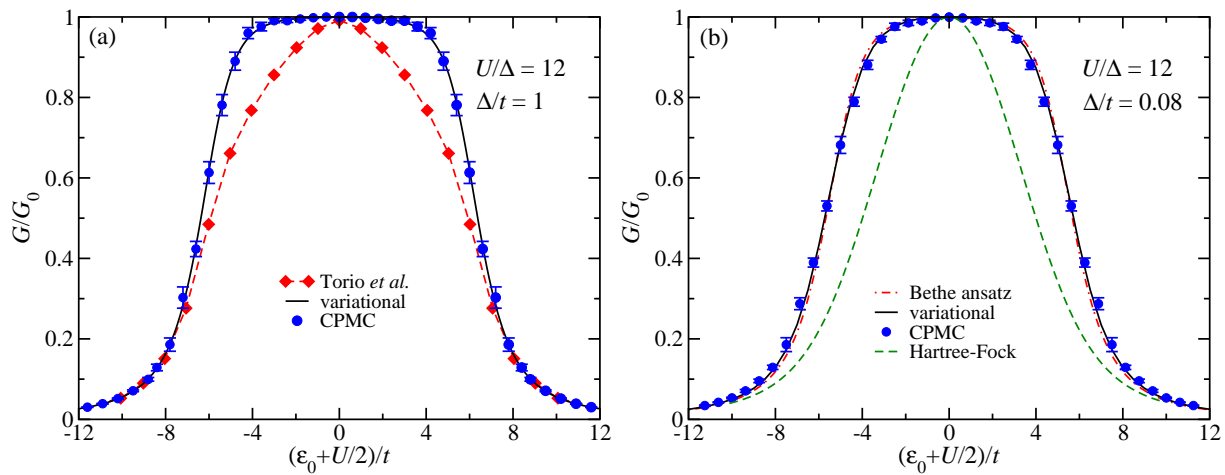

\begin{center}\includegraphics[%
  bb=210bp 0bp 610bp 652bp,
  width=5cm,
  keepaspectratio]{Fig1a.eps} \hspace{3cm}\includegraphics[%
  bb=210bp 0bp 610bp 652bp,
  width=5cm,
  keepaspectratio]{Fig1b.eps}\end{center}

\caption{\label{cap:Fig1}(a) In $2t\ll U$ regime the conductance of THCP
(diamonds) compared to results obtained from the two-point sine conductance
formula,\cite{rejec03} using variational wave function of Gunnarsson
and Schönhammer (full line)\cite{schonhammer85} and CPMC method (bullets).\cite{bonca04}
(b) Conductance for $2t\gg U$ obtained with exact Bethe ansatz (dashed-dotted)\cite{wiegman80}
and Hartree-Fock (dashed). In numerical calculation $t=1.04U$ was
used for the case of variational wave function (full line)\cite{rejec03,schonhammer85}
and CPMC (bullets).\cite{bonca04}}
\end{figure}

\narrowtext

In summary, the method used by THCP seems to be promising for the
calculation of conductance of systems where the interaction is very
weak. However, in the limit of strong electron-electron interaction
it fails to reproduce many-body effects as is, e.g., the Kondo physics
in transport through a quantum dot.

Authors acknowledge the support of the Ministry of Education, Science
and Sport of Slovenia under grant Pl-0044.


\begin{thebibliography}{10}
\bibitem{parameters}All model parameters throughout this Comment are defined as in Ref.~\onlinecite{torio02}.
\end{thebibliography}
\end{document}